# Measurement of average cross sections and isomeric ratios for $^{nat}$Re(γ,xn) photonuclear reactions at the end-point bremsstrahlung energies of 30 MeV and 40 MeV

*A. Alikhanyan National Science Laboratory (YerPhI), 2 Alikhanian Br.Street,*
*0036 Yerevan, Armenia*

A.E. Avetisyan, R.V. Avetisyan*, A.G. Barseghyan, R.K. Dallakyan, Yu.A. Gharibyan,

A.V. Gyurjinyan, I.A. Kerobyan, H.A. Mkrtchyan

**Abstract**

The cross sections for (γ,xn) reactions at 30 MeV and 40 MeV bremsstrahlung end-point energies on natural rhenium ($^{nat}$Re) targets have been measured by the activation and the off-line γ-ray spectrometric techniques using a High Purity Germanium detector (HPGe). The measured $^{nat}$Re(γ,xn) reaction cross sections were compared to the theoretically calculated cross sections using TALYS 1.9 and EMPIRE 3.2 computer codes. The measurements allowed the determination of the Isomeric Cross section Ratios (ICR) for the $^{184m,g}$Re and $^{182m,g}$Re isomeric pairs. In addition, we have determined a semi-empirical value for the $^{186m,g}$Re isomeric pair. These results for the 30 MeV and 40 MeV end-point bremsstrahlung energies are obtained for the first time.

**Introduction**

Photonuclear reaction cross section data are of great importance to study the structure of nuclei and mechanisms of nuclear reactions. Researchers are interested in the isomeric yield ratios from various photo-nuclear reactions for a variety of applications [1]. Rhenium is interesting in particular since it is often used in nuclear medicine for radiotherapy and it is an issue in radiation waste solutions. These applications, in particular for radiation shielding designs, the calculations of absorbed dose in the human body during radiotherapy, activation analysis, nuclear waste transmutation, astrophysical nucleosynthesis [2], physics of fusion and fission reactors etc [3]. The bremsstrahlung cross sections for rhenium have been measured by several experiments [4-6], using mainly the monoenergetic or quasi-monoenergetic photon beams, as well as in different energy regions than evaluated in current work. In this work, we have carried out the theoretical calculations and made the measurements with 30 MeV and 40 MeV bremsstrahlung for the first time.

*rave@yerphi.am

The experimental investigations of the ICR, as a function of incident particle energy and reaction channel, could lead useful information about the spin-cutoff parameter and on the level structure of the product nucleus [7]. On the other hand, different theoretical models variously describe nuclear reaction mechanisms. The information about the isomer ratio of residual nuclei is needed to verify the different models used to explain the nuclear reaction mechanism, optimize the production yield and estimation of the impurities of radioisotopes simultaneously produced.

The goal of this work is to measure the flux-weighted average cross sections of the $^{nat}Re(\gamma,xn)^{182m,g;183;184m,g;186}Re$ reactions and ICRs of final isomeric pairs.

**Experimental details**

*Irradiation*

The irradiation of natural rhenium target was performed by using the linear electron accelerator LUE-75 [8] at the AANL, in Yerevan, Armenia. The nuclear characteristic data of the final isotopes are presented in the Table 1.

| Reaction product | Spin | Half life | Decay mode (%) | $E_\gamma$ (keV) | $I_\gamma$ (%) |
|---|---|---|---|---|---|
| $^{182g}Re$ | $7^+$ | 64.2 h | 100 ($\beta+$) | 351.07 | 10.3 |
| $^{182m}Re$ | $2^+$ | 14.14 h | 100 ($\beta+$) | 470.26 | 2.02 |
| $^{183}Re$ | $5/2^+$ | 70 d | 100 (EC) | 162.33 | 25.1 |
| $^{184g}Re$ | $3^-$ | 35.4 d | 100 ($\beta+$) | 792.07 / 903.28 | 17.22 / 38.1 |
| $^{184m}Re$ | $8^+$ | 169 d | 75.4 (IT) / 24.6 ($\beta+$) | 216.54 | 9.5 |
| $^{186g}Re$ | $1^-$ | 3.72 d | 93.1 ($\beta^-$) / 6.9 (EC) | 137.2 | 9.47 |
| $^{186m}Re$ | $8^+$ | $2 \times 10^5$ y | 90 (IT) / 10 ($\beta^-$) | 40.4 / 59 | 5 / 17.8 |

Table 1. Decay data of measured reaction products

The linear accelerator was operated at 30 MeV and 40 MeV energies. The electron beam diameter was measured and determined to be 20 mm [8]. The targets were irradiated for 2 hrs and 1 h with 0.45 µA and 1.1 µA beam currents, respectively. The electron beam current was stable during the irradiation time. Thus, the current produced constant flux of photons throughout the irradiation.

An hour after the irradiation, rhenium and copper foils were removed from the target module and transferred to the special room for further spectrometric measurements. The gamma-spectroscopy of irradiated targets was performed using an ORTEC HPGe detector with an analyzer and the MAESTRO software. Several measurements were carried out on the activated samples and the averages are presented as final results.

*Target preparation*

The bremsstrahlung photon beam was generated when the electron beam hits 2 mm tantalum (Ta) converter, placed in front of the electron beam exit [9]. In order to measure the $^{nat}Re(\gamma,xn)$ $^{182m,g;183;184m,g;186}Re$ reactions cross-sections the bremsstrahlung flux was monitored by the $^{nat}Cu(\gamma,xn)^{64}Cu$ reaction. Natural copper consist of two $^{63}Cu$ (69.17%) and $^{65}Cu$ (30.83%) stable isotopes. The $^{64}Cu$ isotope occurs only on $^{65}Cu$ isotope by $(\gamma,n)$ reaction, therefore in data analysis the $^{65}Cu(\gamma,n)^{64}Cu$ as a monitor reaction was used.

The target-stack of Cu-Re was installed in the target module immediately after the tantalum converter. The electron beam shape was elliptical with a 2 cm major and a 1.4 cm minor axis.

Natural rhenium targets (purity 99.99 %) supplied by ADVENT Research Materials Ltd. (Oxford, England), were irradiated [10]. The sizes of all used targets and monitor foils are presented in Table 2.

| Energy (MeV) | Nuclide | Mass (g) | Sizes (cm) | Thickness (cm) |
|---|---|---|---|---|
| 30 | $^{nat}Cu$ | 0.224 | 2.5x2.5 | 0.004 |
| | $^{nat}Re$ | 3.022 | 2.3x2.5 | 0.025 |
| 40 | $^{nat}Cu$ | 0.224 | 2.5x2.5 | 0.004 |
| | $^{nat}Re$ | 3.284 | 2.5x2.5 | 0.025 |

Table 2. Details of the irradiated targets along with the copper control foils for monitoring the reactions

**Data analysis**

*Measurement technique*

Gamma-ray spectra of irradiated targets were measured off-line by ORTEC HPGe detector. The detector calibration by standard radioactive sources in different distances from the surface of detector was published earlier [11]. The samples and copper monitors were measured separately and were counted at fixed position from the active surface of the detector. The produced nuclides were identified from knowledge of the target nuclide, the decay data and gamma-ray spectra. Data are taken from National Nuclear Data Center (NNDC) [12] and listed in the Table 1.

The cross sections of interest of reactions were calculated by the following equation [9]:

$$\sigma = \frac{\Delta N \, \lambda}{\varepsilon \eta k N_\gamma \, N_{nucl}(1 - e^{-\lambda t_1})e^{-\lambda t_2}(1 - e^{-\lambda t_3})} \quad (1)$$

where $\Delta N$ is the number of events under the photopeak; $\lambda$ is the decay constant of the residual radioisotope; $N\gamma$ is the beam intensity; $N_{nucl}$ is the number of target nuclei; $k$ is the coefficient of gamma absorption in the target, in air, and in the detector lid; $\varepsilon$ is the HPGe detector efficiency; $\eta$ is the partial intensity of the product gamma line; $t_1$ is the irradiation time; $t_2$ is the time between the end of irradiation and the beginning of measurement; $t_3$ is the duration of the measurement.

*Photon Flux calculation*

The experimental bremsstrahlung $N_\gamma$ photon flux was determined by using the $\Delta N$ photo-peak area based on the activity of 511 keV 35.2% γ-line of $^{64}Cu$ from the $^{65}Cu(\gamma,n)$ monitor reaction. It was calculated by the following equation:

$$N_\gamma = \frac{\Delta N \, \lambda}{<\sigma> \varepsilon \eta k \, N_{nucl}(1 - e^{-\lambda t_1})e^{-\lambda t_2}(1 - e^{-\lambda t_3})} \quad (2)$$

where all the terms have the same meaning as in equation (1).

The experimental photo-nuclear cross section of $^{65}Cu(\gamma,n)^{64}Cu$ reaction for 30 MeV and 40 MeV bremsstrahlung end-point energies have been taken from [13].

The theoretical flux-weighted average cross sections of $^{nat}Re(\gamma,xn)$ $^{182m,g;183;184m,g;186}Re$ reactions have been calculated by the following relation [14]:

$$<\sigma> = \frac{\sum \sigma \varphi}{\sum \varphi} \quad (3)$$

where $\sigma$ is the studied reaction cross section at $E_\gamma$ energy and $\varphi$ is the photon flux at the same $E_\gamma$ energy. The theoretical photo-nuclear cross sections for the particular energy were calculated by TALYS 1.9 [15] (default mode) and EMPIRE 3.2 [16] (exciton mode) nuclear codes. The photon flux distribution for bremsstrahlung end-point energies of 30 MeV and 40 MeV were calculated by the GEANT4 code [17]. The simulations have been done by taking into account the beam collimation and the realistic sizes of tantalum converter. The photon flux distribution as function of photon energy generated for one million electrons is presented in Fig.1.

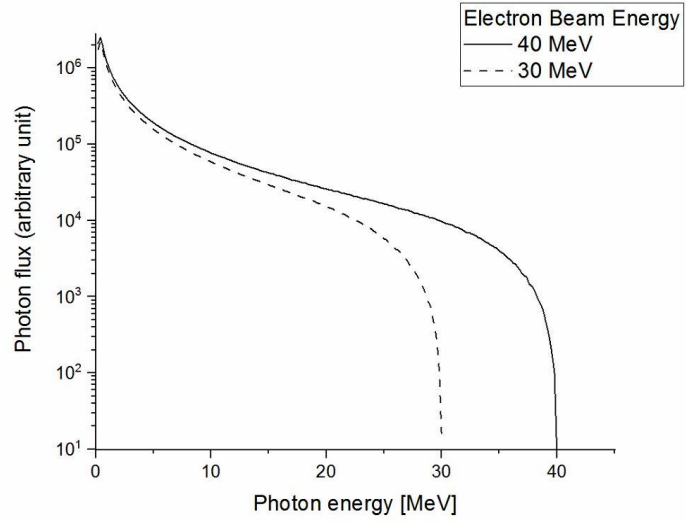

Fig. 1. Bremsstrahlung spectrum from the 30 MeV and 40 MeV electron beam calculated by GEANT4

Since the $^{65}$Cu(γ,n)$^{64}$Cu and $^{nat}$Re(γ,xn)$^{186,184,183,182}$Re reactions thresholds are different, the photon flux obtained from the $^{65}$Cu(γ,n)$^{64}$Cu reaction by equation (2) has to be modified taking into account the threshold energy of each studied reaction. For $^{187}$Re(γ,xn)$^{186,184,183,182}$Re reactions, the average flux obtained from the $^{65}$Cu(γ,n)$^{64}$Cu reaction is multiplied by a different conversion $C_x$ factor, based on the threshold ratio values. The conversion $C_x$ factors at the bremsstrahlung end-point energies of 30 MeV and 40 MeV were calculated taking into account the equation (4) [18]:

$$C_x = \int_{E_{thr,x}^{Re}}^{E_e} \Phi(E)dE \bigg/ \int_{E_{thr}^{Cu}}^{E_e} \Phi(E)dE \qquad (4)$$

where $E_{thr,x}^{Re}$ and $E_{thr}^{Cu}$ are the threshold energies of $^{nat}$Re(γ,xn)$^{182,183,184,186}$Re reactions and $^{65}$Cu(γ,n)$^{64}$Cu monitor reaction, respectively. The bremsstrahlung spectrum, i.e the photon flux distribution $\Phi(E)$ as a function of photon energy were simulated by GEANT4 (see Fig. 1) and the corresponding threshold values from Table 3 to calculate the conversion $C_x$ factors were used. In order to determine the number of real photons for the reactions under study, the flux value obtained from (2) was multiplied by the corresponding $C_x$ correction factor obtained from (4). The values of $C_x$ coefficients with reactions thresholds are presented in Table 3.

**Results and discussions**

*Cross sections*

The flux-weighted average cross sections for the $^{nat}Re(\gamma,xn)^{182m,g;183;184m,g;186g}Re$ reactions received using photon flux from experimental data [13] and from theoretical calculations by TALYS 1.9 and EMPIRE 3.2 are presented in Table 3.

| Isotope | Reaction thresholds (MeV) | Bremsstrahlung end-point energy (MeV) | Flux conversion ratios ($C_x$) | Average cross section (mb) | | |
|---|---|---|---|---|---|---|
| | | | | Experimental | TALYS 1.9 | EMPIRE 3.2 |
| $^{182g}Re$ | 22.8 | 30 | 0.079 | 0.45±0.089 | 0.64 | - |
| | | 40 | 0.224 | 0.93±0.001 | 1.34 | - |
| $^{182m}Re$ | 22.6 | 30 | 0.079 | 2.27±0.209 | 1.2 | - |
| | | 40 | 0.224 | 2.26±0.113 | 1.84 | - |
| $^{183}Re$ | 14.2 | 30 | 0.518 | 16.56±0.244 | 30.3 | 40.02 |
| | | 40 | 0.623 | 15.62±0.037 | 24.2 | 31.71 |
| $^{184g}Re$ | 7.8 | 30 | 1.409 | 29±0.134 | 48.21 | 43.72 |
| | | 40 | 1.305 | 24.06±0.127 | 42.94 | 39.13 |
| $^{184m}Re$ | 8 | 30 | 1.366 | 0.78±0.048 | 1.26 | 0.47 |
| | | 40 | 1.273 | 0.76±0.025 | 1.3 | 0.5 |
| $^{186g}Re$ | 7.4 | 30 | 1.503 | 93.37±0.133 | 126.96 | 100.03 |
| | | 40 | 1.374 | 92.12±0.131 | 110.45 | 87.01 |

Table 3. The flux-weighted average cross sections for $^{nat}Re(\gamma,xn)^{182m,g;183;184m,g;186g}Re$ reactions on the base of experimental data [13] and calculations by TALYS 1.9 and EMPIRE 3.2

As can be seen from the Table 3, the cross section values based on the experimental monitor reaction data from [13] for all isotopes (except $^{182m}Re$) are lower than the results based on theoretical calculations by TALYS 1.9 and EMPIRE 3.2 codes. Along with this, from mentioned two theoretical calculations, the results from EMPIRE 3.2 are closer to the experimental results.

*ICRs of . $^{182m,g}Re$, $^{184m,g}Re$ and $^{186m,g}Re$*

In current work the ICR is defined as the ratio of isomeric cross section to the total cross section ($\sigma_m/\sigma_{tot}$). For the $^{182m,g;184m,g}Re$ isomeric pairs both theoretically and experimentally ICRs were determined. Experimental values for isomeric ratios were obtained using measured data on average $\sigma_m$ and $\sigma_{tot}$ cross sections. For the theoretical estimation of isomeric ratios, average cross sections were calculated by TALYS 1.9 and EMPIRE 3.2 nuclear codes (see Equation (3)). The bremsstrahlung end-point energies of 30 MeV and 40 MeV gamma-ray spectra simulated by GEANT4 were used to calculate the average cross sections.

The direct measurement of ICR for $^{186m,g}$Re isomeric pair is impossible due to the long half-life of the $^{186m}$Re isomeric state ($2 \times 10^5$ y). The half-life of the $^{186g}$Re ground state is short (3.72 d), therefore the experimental average cross section measurement is possible. In order to evaluate the ICR for $^{186m,g}$Re isomeric pair the semi-empirical method is used which consists of the combination of theoretical and experimental data. In order to obtain the theoretically calculated total cross section for the formation of $^{186}$Re, the TALYS 1.9 and EMPIRE 3.2 nuclear codes were applied.

The comparison of experimental and theoretical ICRs for the $^{182m,g}$Re and $^{184m,g}$Re isomeric pairs as well as the semi-empirical results for the $^{186m,g}$Re isomeric pair are presented in Table 4 and Table 5, respectively.

| Isomeric pair | Bremsstrahlung end-point energy (MeV) | Isomeric Cross Section Ratio | | |
|---|---|---|---|---|
| | | Experimental | TALYS 1.9 | EMPIRE 3.2 |
| $^{182m,g}$Re | 30 | 0.835±0.03 | 0.65 | - |
| | 40 | 0.709±0.01 | 0.58 | - |
| $^{184m,g}$Re | 30 | 0.026±0.002 | 0.025 | 0.011 |
| | 40 | 0.031±0.001 | 0.029 | 0.013 |

Table 4. The experimental and theoretical ICRs for $^{182m,g}$Re and $^{184m,g}$Re isomeric pairs

| Isomeric Pair | Bremsstrahlung end-point energy (MeV) | Isomeric Cross Section Ratio | | | |
|---|---|---|---|---|---|
| | | Semi-empirical | | Theoretical | |
| | | TALYS 1.9 | EMPIRE 3.2 | TALYS 1.9 | EMPIRE 3.2 |
| $^{186m,g}$Re | 30 | 0.03 | 0.018 | 0.023 | 0.017 |
| | 40 | 0.029 | 0.0166 | 0.024 | 0.0176 |

Table 5. The semi-empirical and theoretical ICRs for $^{186m,g}$Re isomeric pairs

The ICR data for the $^{182m,g}$Re and $^{184m,g}$Re isomeric pairs are obtained for the first time. As can be seen from Tables 4 and 5 the ICRs calculated by EMPIRE 3.2 are lower than the results from TALYS 1.9 for all studied reactions. In general, the results from TALYS 1.9 are closer to the experimental results.

**Conclusion**

The flux-weighted average cross sections for the $^{nat}Re(\gamma,xn)^{186,184,183,182}Re$ reactions at the bremsstrahlung end-point energies of 30 MeV and 40 MeV were determined by activation and the off-line γ-ray spectrometric technique. Theoretically the reactions cross sections were also calculated by the TALYS 1.9 and EMPIRE 3.2 codes and the results found to be in general agreement with each other. Because of the lack of experimental data in literature corresponding to this energy region, there is no comparison of different works with the current calculations and measurements. For more comprehensive analysis of the theoretical calculations and experimental measurements, further investigations will needed to be done in the more wide energy region.

*Acknowledgments*

The authors are thankful to the staff of the electron linear accelerator machine group at the A. Alikhanyan National Science Laboratory, for the stable operation of the accelerator and for their support during the experiment.

This work was supported by the RA MESCS State Committee of Science, in frames of the research project № SCS 18T-1C297.

**References**

[1] M. Tatari, G. Kim, H. Naik, K. Kim, S. C. Yang, M. Zaman, S. G. Shin, Y.U. Kye, M.H. Cho, Nuclear Instruments and Methods in Physics Research B 344 (2015) 76–82.

[2] D. D. Clayton, Astronomy with Radioactivities, pp 25-79, 2010.

[3] R. Crasta, H. Naik, S. V. Suryanarayana, S. Ganesh1, P. M. Prajapati, M. Kumar, T. N. Nathaniel, V. T. Nimje, K. C. Mittal, A. Goswami1, Radiochim. Acta 101, 541–546 (2013).

[4] A. Veyssiere, H. Beil, R. Bergere, P. Carlos, A. Lepretre, A. De Miniac, Journal de Physique Lettres Vol.36, p.267, 1975.

[5] T. Shizuma, H. Utsunomiya, P. Mohr, T. Hayakawa, S. Goko, A. Makinaga, H. Akimune, T. Yamagata, M. Ohta, H. Ohgaki, Y.-W. Lui, H. Toyokawa, A. Uritani, S. Goriely, Physical Review, Part C, Nuclear Physics Vol.72, p.025808, 2005.

[6] A. M. Goryachev, G. N. Zalesnyy, S. F. Semenko, B. A. Tulupov, Yadernaya Fizika, Vol.17, p.463, 1973.

[7] R.V. Avetisyan, Journal of Contemporary Physics 54, 338–344 (2019).

[8] A. Sirunyan, A. Hakobyan, G. Ayvazyan, A. Babayan, H. Vardanyan, G. Zohrabyana, K. Davtyan, H. Torosyan, A. Papyan, Journal of Contemporary Physics 53, 271-278, (2018).

[9] A.S. Danagulyan, G.H. Hovhannisyan, T.M. Bakhshiyan, A. E. Avetisyan, I. A. Kerobyan, R. K. Dallakyan, Phys. Atom. Nuclei 78, 447–452 (2015).

[10] ADVENT research materials, https://www.advent-rm.com/.

[11] R. K. Dallakyan, Armenian Journal of Physics, 2013, vol. 6, issue 1, pp. 45-50.

[12] National Nuclear Data Center, NuDat 2.8, https://www.nndc.bnl.gov/nudat2/.

[13] K. Masumoto, T. Kato, N. Suzuki, Nuclear Instruments and Methods 157, 567-577 (1978).

[14] H. Naik, G.N. Kim, R. Schwengner, K. Kim, M. Zaman, M. Tatari, M. Sahid, S.C. Yang, R. John, R. Massarczyk, A. Junghans, S.G. Shin, Y. Key, A. Wagner, M.W. Lee, A. Goswami, M.-H. Cho, Nuclear Physics A 916 (2013) 168–182.

[15] A. Koning, S. Hilaire, S. Goriely, TALYS 1.9 nuclear reaction program, 2017.

[16] M. Herman, R. Capote, M. Sin, A. Trkov et al., EMPIRE-3.2 Rivoli modular system for nuclear reaction calculations and nuclear data evaluation, 2013.

[17] S. Agostinelli, et al., GEANT4 - A simulation toolkit. Nuclear Instruments and Methods in Physics Research Section A, 2003, vol. 506, no. 3, pp. 250–303.

[18] H. Naik, G. Kim, K. Kim, M. Zaman, A. Goswami, M. WooLee, S-C. Yang, Y-O. Lee, S-G. Shin, M-H. Cho, Nuclear Physics A 948 (2016) 28.